\documentclass[a4paper,11pt]{article}
\usepackage{jheppub}
\usepackage[T1]{fontenc}
\usepackage[utf8]{inputenc}
\usepackage[american]{babel}
\usepackage{csquotes}
\usepackage{subcaption}

\usepackage{float}
 
\usepackage{lipsum}
\usepackage{graphicx}
\usepackage{dcolumn} 
\usepackage{tabularx}
\usepackage{bm} 
\usepackage{amssymb}
\usepackage{amsmath}

\usepackage{booktabs}
\usepackage{multirow}
\usepackage{makecell}
\usepackage[colorlinks=true]{hyperref}

\title{\boldmath Signal identification with Kalman Filter towards background-free neutrinoless double beta decay searches in gaseous detectors}

\author[a]{Tao Li,}
\author[b,c,1]{Shaobo Wang,\note{Corresponding author.}}
\author[a]{Yu Chen,}
\author[b,1]{Ke Han,}
\author[b]{Heng Lin,}
\author[b]{Kaixiang Ni,}
\author[a]{Wei Wang,}
\author[c]{Yiliu Xu,}
\author[c]{and An'ni Zou}

\affiliation[a]{School of Physics, Sun Yat-Sen University, Guangzhou, {\rm 510215}, China}
\affiliation[b]{INPAC; Shanghai Laboratory for Particle Physics and Cosmology; 
 Key Laboratory for Particle Astrophysics and Cosmology (MOE), \\
School of Physics and Astronomy, Shanghai Jiao Tong University, Shanghai {\rm 200240}, China}
\affiliation[c]{SPEIT~(SJTU-ParisTech Elite Institute of Technology), Shanghai Jiao Tong University, Shanghai, {\rm 200240}, China}

\emailAdd{shaobo.wang@sjtu.edu.cn}
\emailAdd{ke.han@sjtu.edu.cn}

\date{\today}
\abstract{
Particle tracks and differential energy loss measured in high pressure gaseous detectors can be exploited for event identification in neutrinoless double beta decay~($0\nu \beta \beta$) searches.
We develop a new method based on Kalman Filter in a Bayesian formalism (KFB) to reconstruct meandering tracks of MeV-scale electrons.
With simulation data, we compare the signal and background discrimination power of the KFB method assuming different detector granularities and energy resolutions.
Typical background from $^{232}$Th and $^{238}$U decay chains can be suppressed by another order of magnitude than that in published literatures, approaching the background-free regime. 
For the proposed PandaX-III experiment, the $0\nu \beta \beta$ search half-life sensitivity at the 90\% confidence level would reach $2.7 \times 10^{26}$~yr with 5-year live time, a factor of 2.7 improvement over the initial design target.
}

\begin{document}

\maketitle
\flushbottom

\section{Introduction}
\label{sec:Introduction}

Neutrinoless double beta decay ($0\nu\beta\beta$) is a hypothetical weak decay process that would confirm the Majorana nature of neutrinos and provide a direct evidence of lepton number violation~\cite{Majorana:1937vz, Avignone:2007fu}. 
Experimental search for $0\nu\beta\beta$ has been an active frontier in particle and nuclear physics~\cite{Dolinski:2019nrj}. 
The Standard-Model-allowed two neutrino double beta decay ($2\nu\beta\beta$) has been observed in more than ten isotopes.
Searches for the $0\nu\beta\beta$ utilize those isotopes, including $^{136}$Xe, $^{76}$Ge, and $^{130}$Te.
The current lower half-life limits for $0\nu\beta\beta$ of the three isotopes are $1.07 \times 10^{26}$~yr, $1.8 \times 10^{26}$~yr, and $3.2 \times 10^{25}$~yr (90\% confidence level, or C.L.), established by KamLAND-Zen, GERDA, and CUORE experiments respectively~\cite{KamLAND-Zen:2016pfg, Agostini:2020xta, Adams:2019jhp}.
Most $0\nu\beta\beta$ experiments identify possible signals by event excess around the decay Q-value in the summed electron energy spectrum.

\begin{figure}[hb] 
  \centering 
  \includegraphics[width=.9\textwidth]{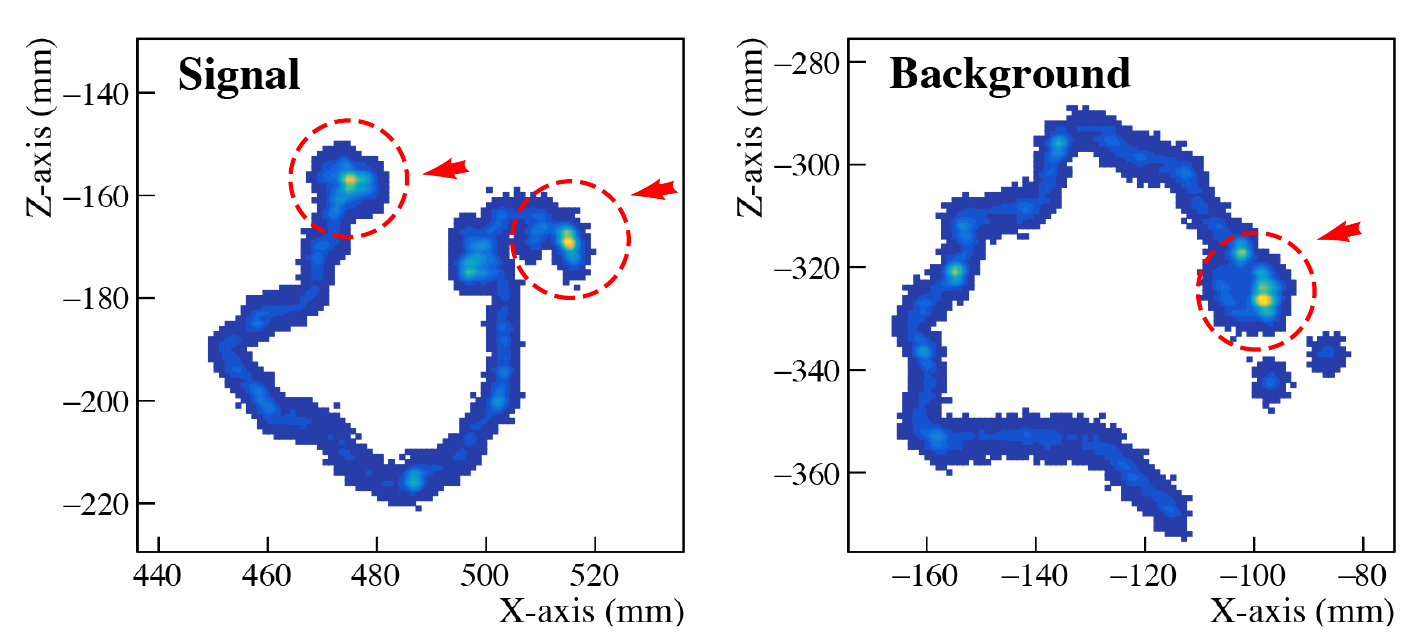}
    \caption{Projections in a 2-dimensional plane of a $0\nu\beta\beta$ signal (left) and a background (right) event in a gaseous xenon TPC with 10 bar pressure and 1~mm pixels readout.
    The red dotted circles indicate the BBs.
    For the $0\nu\beta\beta$ signal event two BBs can be seen at the ends while only one BB for the background event.
    } 
    \label{fig:track_example} 
\end{figure}

Xenon-based high pressure gaseous Time Projection Chambers (TPCs) can utilize topological features of event tracks and differential energy loss to identify possible $^{136}$Xe $0\nu\beta\beta$ signals.
Two emitted electrons carry the Q-value of 2458~keV and may travel $\mathcal{O}$ (20~cm) along meandering tracks in 10~bar xenon gas.
Signal identification with tracks in a $^{136}$Xe gaseous TPC has been exploited early on by the Gotthard experiment~\cite{Luscher:1998sd}.
More recently, the NEXT experiment~\cite{Rogers:2018lle} demonstrated the effective track reconstruction with electroluminescence amplification signals in their prototype TPC~\cite{Ferrario:2015kta}.
Efficiencies of signal and background identification in the proposed PandaX-III gaseous TPC~\cite{Chen:2016qcd} have been studied with simulation data~\cite{Qiao:2018edn, Galan:2019ake}.
The aforementioned experiments utilize the prominent Bragg Blob (BB) feature of event tracks. 
Each end of a $0\nu\beta\beta$ track has a BB, in which rapid energy loss in unit volume happens because of increased differential energy loss (Bragg peak) and larger scattering angles right before an electron stops.  
However, only one BB are on the ends of the track to the background.
Fig.~\ref{fig:track_example} shows typical tracks of $0\nu\beta\beta$ signal and background events from simulation in  gaseous xenon TPC, where the track of $0\nu\beta\beta$ has two BBs while that of background from the $^{238}$U decay chain has one.

Our study aims to reconstruct the meandering tracks with Kalman Filter in a Bayesian formalism (KFB)~\cite{doi:10.1002/acs.2369,Frosini:2017ftq} and extract more signal-identifying features besides BB~\cite{Luscher:1998sd,Ferrario:2015kta}.
For example, differential energy loss along tracks and particle momentum by the end of the tracks can be calculated and used for the signal and background identification.
When combined with parameters such as track energies, we demonstrate that almost all the background in the region of interest (ROI) of $0\nu\beta\beta$ search can be rejected and the search sensitivity further improved in a typical low-background gaseous TPC.

The paper is organized as follows: 
In Section~\ref{sec:Simulation}, the generation of simulation data is presented. 
Five detector configurations are studied to compare the effect of the readout schemes and detector energy resolutions.
In Section~\ref{sec:Method}, we show the preprocessing and the method of track reconstruction with KFB in details. 
Then topological parameters are extracted for signal identification in Section~\ref{sec:Parameters}.
In Section~\ref{sec:result}, the signal and background discrimination power and the improvement for $0\nu\beta\beta$ search sensitivity of a proposed detector are presented.
At last, we discuss the potential improvement and broader applications of KFB.

\section{Simulation}
\label{sec:Simulation}

To demonstrate the performance of the KFB approach, we construct a conceptual high pressure gaseous TPC with the Geant4 simulation framework~\cite{Agostinelli:2002hh}.
The signal events are the $0\nu\beta\beta$ of $^{136}$Xe.
Background events from the decay chains of $^{232}$Th and $^{238}$U are considered in the simulation. 
To explore the impact of event discrimination power of different readout schemes and detector energy resolution, five detector configurations are considered.

\subsection{Geometry and event generation}

The detector geometry is similar to the PandaX-III TPC as outlined in Ref.~\cite{Wang:2020owr}. 
The active volume (AV) is 1.6~m in diameter and 1.2~m high, which contains approximately 140~kg of xenon gas (with 90\% $^{136}$Xe) at 10 bar.
The cathode and the readout plane are placed at the two bases of the cylindrical AV respectively. 
Outside of the AV, we construct an acrylic field cage, copper shielding liner, stainless steel vessel, lead shielding, and high-density polyethylene shielding in sequence.
The dimensions of each component are identical to Ref.~\cite{Xie:2020xmd}.
The $0\nu\beta\beta$ signals are produced with the Decay0 package~\cite{Ponkratenko:2000um}, which gives the energy distributions of two emitted electrons.
Background events from the decay chains of $^{232}$Th and $^{238}$U in the detector components and shielding layers are considered.

The detector response of TPC, including electron diffusion, readout schemes, and energy resolution, is simulated in the REST framework~\cite{Galan:2019ake}.
While drifting to the readout plane, ionization electrons diffuse transversely and longitudinally which results in broadening of the tracks.
The transverse (longitudinal) diffusion coefficient is set to be $1.0\,(1.5)\times10^{-2}$~cm$^{1/2}$, assuming xenon is mixed with a quencher gas such as Trimethylamine to have reduced diffusions.
Detector response blurs event tracks spatially and energetically, which decreases the discrimination power between signal and background events.
Hence, it is critical to reconstruct the tracks accurately for effective background suppression.

\subsection{Detector configurations}

We have performed our studies under five different configurations by varying the readout schemes and detector energy resolutions.

In the first two high granularity configurations, the detector's readout plane is fully instrumented with 1~mm~$\times$~1~mm pixels.
An energy resolution of 3\% (Full Width at Half Maximum, FWHM) at the Q-value is assumed for the first one and 6\% for the second. 
We compare such detectors with configurations of degraded spatial granularity of 3~mm~$\times$~3~mm and energy resolution of 3\% and 1\% respectively in the third and fourth configurations.

The last configuration replicates the PandaX-III detector specifications.
The readout plane is covered with 52 pieces of 20~cm~$\times$~20~cm readout modules, each of which is equipped with 3-mm-wide readout strips.
Each strip reads out signals from 64 connected pixels in horizontal or vertical directions.
The setup significantly reduces the number of readout channels needed, but does introduce ambiguity in track reconstruction.
It is also worth noting that efficiency loss is considered because the TPC's AV is not 100\% monitored by readout modules in this configuration. 
The energy resolution is assumed to be 3\%.

Later we will refer the configurations as \emph{(1~mm, 3\%)}, \emph{(1~mm, 6\%)}, \emph{(3~mm, 3\%)}, \emph{(3~mm, 1\%)}, and \emph{(3~mm strip, 3\%)} respectively.
Once reaching the readout plane, ionization electrons register \emph{charge signal hits} (or \emph{hits} for short) in pixels or strips.
The amplitude and timing of hits contain all the information that can be collected in a physical detector.
The output hit data from our simulation would mimic detector data.

\section{Track reconstruction with KFB}
\label{sec:Method}

Kalman Filter is widely used in particle physics experiment~\cite{Adam:2003kg,Piacquadio:2008zza,Chatterjee:2014vta}.
It is used as an optimal estimator for track reconstruction which combines information from physical model prediction and measurement data.
Besides track fitting, it can also be used for optimum extrapolation, 
identification of pseudo-points, and error adjustment~\cite{Fruhwirth:178627}.
The KFB approach combines Kalman filtering with the Bayesian formalism and can estimate the noise covariance matrix of model prediction and measurement.
Focusing on the signal identification in $0\nu\beta\beta$ search, we develop a new method based on KFB to reconstruct the meandering tracks of MeV-scale electrons.
The hits of the simulation data are grouped into principal and subordinate tracks firstly in the preprocessing steps, then the principal track is reconstructed with KFB.

\subsection{Preprocessing}
The $0\nu\beta\beta$ signal and background events may generate more than one track in gaseous TPC.
We focus on the principal track which carries the most deposited energy and contains the most information.
The other shorter tracks (if exist) are named subordinate track(s).
The simulation data are preprocessed in two phases before reconstruction with the KFB approach.

\begin{figure}[tb]
\begin{subfigure}{\textwidth}
  \centering
  \includegraphics[width=0.74\linewidth]{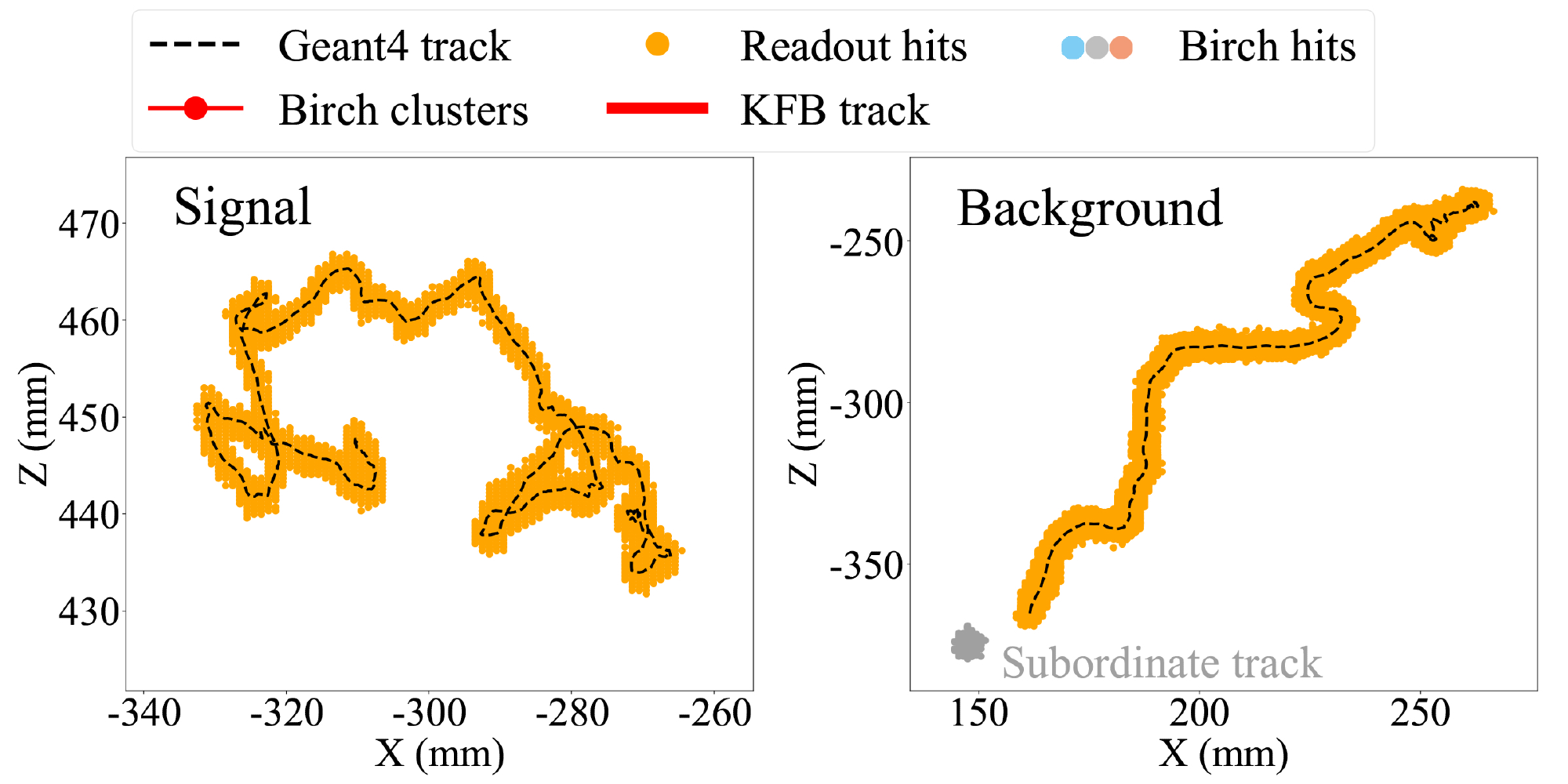}  
  \caption{Identification of the principal track}
  \label{fig:trajectory_0}
\end{subfigure}
\begin{subfigure}{\textwidth}
  \centering
  \includegraphics[width=0.74\linewidth]{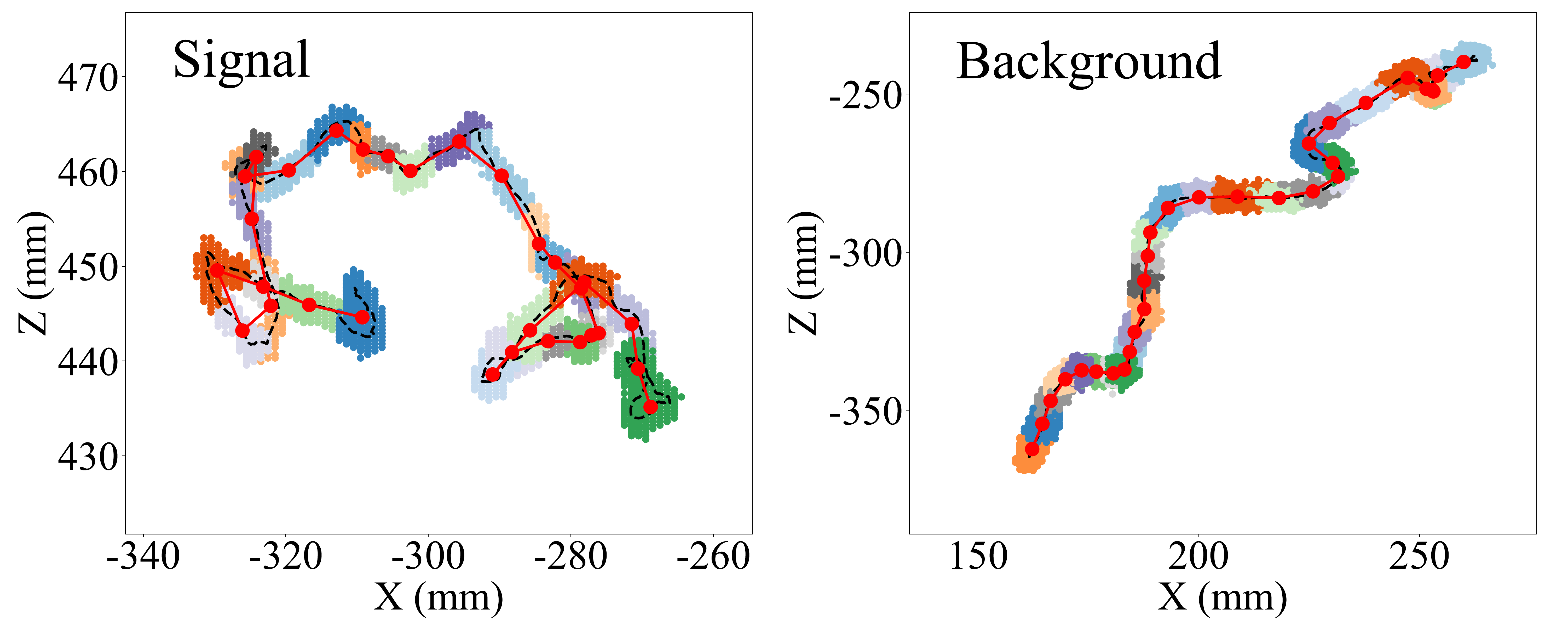}  
  \caption{Rough reconstruction}
  \label{fig:trajectory_1}
\end{subfigure}
\begin{subfigure}{\textwidth}
  \centering
  \includegraphics[width=0.74\linewidth]{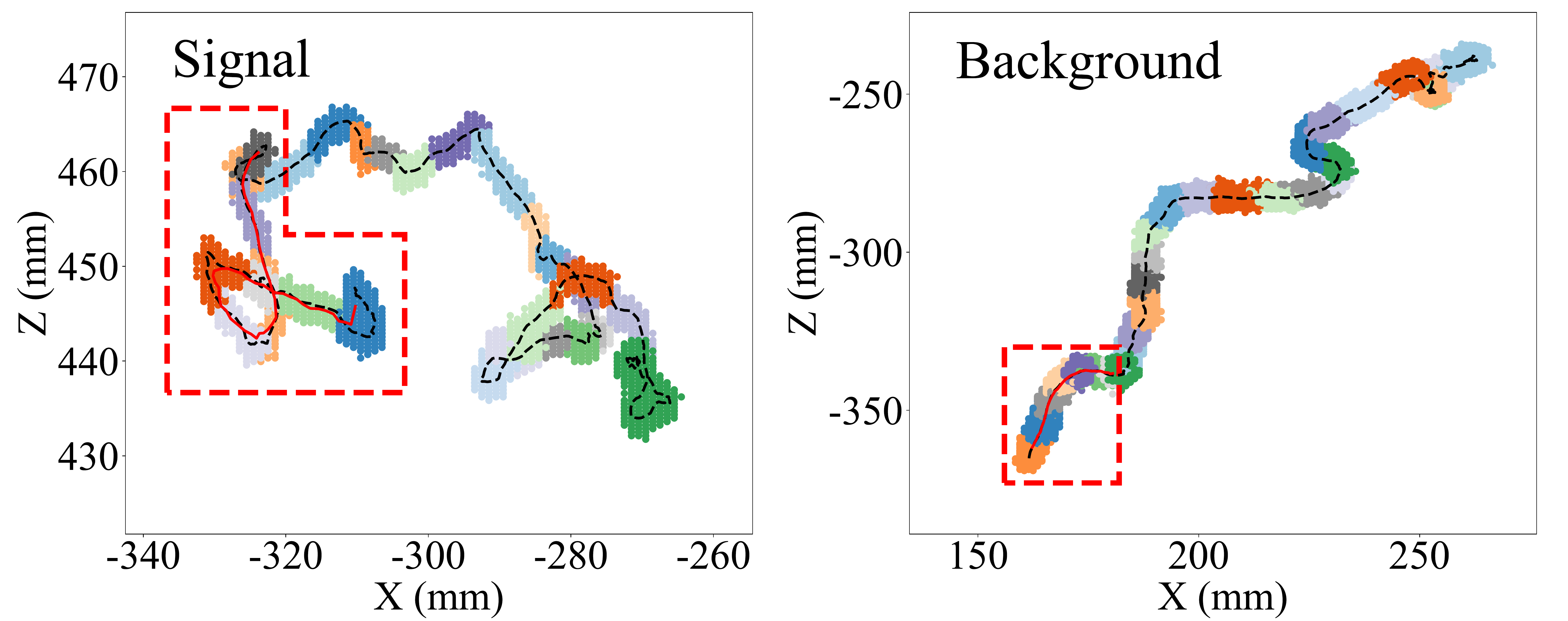}  
  \caption{Reconstruction with KFB}
  \label{fig:trajectory_2}
\end{subfigure}
\caption{Track reconstruction procedures of simulated $0\nu \beta \beta$ signal (left side) and background (right side) events, including (a) identification of the principal track, (b) rough reconstruction, and (c) reconstruction with KFB, respectively.
The tracks are projected in the XZ plane, where X is one of the transverse directions and Z is the longitudinal direction along the field lines in a TPC. 
The MC-truth track from Geant4 (black dashed lines), simulated readout hits (orange points), Birch clustering hits (multi-colored points), BCs (red dot dash lines), and the reconstructed tracks by the ends (red lines) are labeled. 
We filter on both sides of the track, but one of the filter result is drawn for illustration.
KFB stops at the BCs on each end.
}
\label{fig:track} 
\end{figure}

Firstly, we group hits of an event into a principal track and subordinate track(s) based on the distance among the hits.
The grouping is implemented with a widely-used clustering algorithm called DBSCAN~\cite{10.5555/3001460.3001507}, which clusters discrete hits based on proximity.
The cluster with the most deposited energy would be selected as the principle track.
An example of well-separated principal and subordinate tracks of a background event is shown in Fig.~\ref{fig:trajectory_0}.

Secondly, the principal track is reconstructed roughly.
Hits of the principal track are divided into segments with the Birch clustering algorithm~\cite{10.1145/235968.233324}, shown as clusters of colored dots in Fig.~\ref{fig:trajectory_1}.
The charge-weighted centers of segments, named Birch clusters (BCs), are shown as red dot dash lines in Fig.~\ref{fig:trajectory_1}.  
They are sorted by a modified Ant Colony Optimization algorithm~\cite{dorigo1997ant} with random starting points. 
The algorithm enumerates different connections of BCs and finds the one with the shortest total track length of all enumerations. 
The number of segments is optimized for the balance of sorting quality and computing loads.

\subsection{Kalman filter}

The Kalman filter is then introduced to further refine the reconstruction of the tracks~\cite{10.1115/1.3662552, Fruhwirth:178627}, which keeps on recursion of the covariance to estimate the optimal position of each sampled hit.
We use the classical Kalman filter which describes a linear dynamic system with given process and measurement noise~\cite{Innes:1992ge}.
Two equations, the prediction equation and measurement equation need to be determined firstly.

For the state prediction, we define a state vector $s=(x,\,y,\,z,\,u_x,\,u_y,\,u_z)^T$ at any given hit, including the three dimensional position and unit velocity vector.
Considering the physical process of multiple scattering between two adjacent steps $k-1$ and $k$, the prediction equation can be written as: 
\begin{equation}
s_k = Fs_{k-1}+\omega_k.
\label{eq:kf_prediction}
\end{equation} 
where $F$ is the propagation matrix of uniform rectilinear motion and $\omega_k$ is the process noise.
The step size is the same as readout granularity, noted as $\lambda$.
$\omega_k$ can be described as $\omega_k=( 0,~0,~0, ~G(0,\theta_x), ~G(0,\theta_y), ~G(0,\theta_z) )^T$, where $G$ is the standard Gaussian distribution with a mean value 0, and the standard deviation $\theta_{x,y,z}$ represents the projection of scattering angle $\theta_{space}^{rms}$ in three dimensions.
Then we can rewrite Eq.~\ref{eq:kf_prediction} as
\begin{equation}
    \begin{bmatrix}
      [x]_k \\
      [y]_k \\
      [z]_k \\
      [u_x]_k \\
      [u_y]_k \\
      [u_z]_k \\
    \end{bmatrix}
     = \\
    \begin{bmatrix}
      1 & 0 & 0 & \lambda & 0 & 0 \\
      0 & 1 & 0 & 0 & \lambda & 0 \\
      0 & 0 & 1 & 0 & 0 & \lambda \\
      0 & 0 & 0 & 1 & 0 & 0 \\
      0 & 0 & 0 & 0 & 1 & 0 \\
      0 & 0 & 0 & 0 & 0 & 1 \\
    \end{bmatrix}
    \begin{bmatrix}
      [x]_{k-1} \\
      [y]_{k-1} \\
      [z]_{k-1} \\
      [u_x]_{k-1} \\
      [u_y]_{k-1} \\
      [u_z]_{k-1} \\
    \end{bmatrix}
      + \\
    \begin{bmatrix}
      0 \\
      0 \\
      0 \\
      G(0, [\theta_x]_k) \\
      G(0, [\theta_y]_k) \\
      G(0, [\theta_z]_k) \\
    \end{bmatrix},
\label{eq:kf_prediction_2}
\end{equation}
Furthermore, for an electron with momentum $p$, Coulomb scattering distribution follows Gaussian approximation for the Moli\`ere's formula~\cite{Patrignani:2016xqp}:
\begin{equation}
	\theta_{space}^{rms} = \frac{19.2\,{\rm MeV}}{ p v} \sqrt{\lambda_0}[ 1 + 0.038 \ln\lambda_0],
\label{eq:theta}
\end{equation} 
where $v$ is the particle velocity and $\lambda_0$ is the step size in the unit of radiation length in the medium.
Energy deposition is not considered to simplify our physics model.

The measurement equation is defined as follows.
Measurement data are anchored by BCs.
In between adjacent BCs, the steps $m_k=(x^m,~y^m,~z^m)_k^T$ are linearly interpolated hit points spaced by the detector granularity $\lambda$.
The measurement equation will be presented as
\begin{equation}
m_k = Hx_k+\delta_k,
\label{eq:kf_measurement}
\end{equation} 
where $H$ is the the measurement matrix determined by observation process of the detector. 
The associated measurement noise is $\delta_k=(G(0,~\sigma),~G(0,~\sigma),~G(0,\sigma))_k^T$, where measurement uncertainty $\sigma$ includes contributions from detector spatial resolution and uncertainties introduced in the pre-processing.
Eq.~\ref{eq:kf_measurement} is expanded in the form of a matrix
\begin{equation}
    \begin{bmatrix}
      [x^m]_k \\
      [y^m]_k \\
      [z^m]_k \\
    \end{bmatrix}
     = \\
    \begin{bmatrix}
      1 & 0 & 0 \\
      0 & 1 & 0 \\
      0 & 0 & 1 \\
    \end{bmatrix}
    \begin{bmatrix}
      [x]_{k} \\
      [y]_{k} \\
      [z]_{k} \\
    \end{bmatrix}
      + \\
    \begin{bmatrix}
       G(0, [\sigma_x]_k) \\
       G(0, [\sigma_y]_k) \\
       G(0, [\sigma_z]_k) \\
    \end{bmatrix},
\label{eq:kf_measurement_2}
\end{equation}

Kalman filter fuses the physical model prediction and measurement to optimize the state vector $s_k$.
Based on the least mean-square estimation, a linear weight is determinated to minimize the covariance, defined $C_k$, between $s_k$ and its true value.
$s_k$ and $C_k$ are updated in every step $k$ through iteration.
The iterative formulae of kalman filter can be derived from Eq.~\ref{eq:kf_prediction} and Eq.~\ref{eq:kf_measurement} and more details can be found in~\cite{10.1115/1.3662552}.

Bayesian formalism is introduced in our model of Kalman filter to determine the values of process noise $\omega_k$ and measurement noise $\delta_k$~\cite{doi:10.1002/acs.2369,Frosini:2017ftq}. 
The covariance matrices of two vectors are denoted as $Q$ and $R$, respectively. 
We specify two sets of values $\mathbb{Q}=\{Q_1,\,Q_2,\,...,\,Q_{n_Q}\}$ and $\mathbb{R}=\{R_1,\,R_2,\,...,\,R_{n_R}\}$ representing the samples of possible process noise and measurement noise. 
Every possible pair $[Q_i, R_j]$, where $Q_i \in \mathbb{Q}$ and $R_j\in \mathbb{R}$, will be selected as the input parameters of $Q_k$ and $R_k$ for Eq.~\ref{eq:KFB}.
With all the measurements up to step $k$, $\mathcal{M}_{k}=\left\{m_1,\,m_2,\,...,\,m_{k}\right\}$, the probability $P(Q_i,\,R_j\,|\,\mathcal{M}_k)$ is calculated according to
\begin{equation}
\begin{split}
  P(Q_i,\,R_j\,|\,\mathcal{M}_k) \propto P(m_k\,|\,\mathcal{M}_{k-1},\,Q_i,\,R_j)\\
  P(Q_{i},\,R_{j}\,|\,\mathcal{M}_{k-1}).
\end{split}
\label{eq:KFB}
\end{equation} 
At last, The values $\theta_{space}^{rms}$ and $\sigma$ at the step $k$ are determined by maximizing the above probability  
\begin{equation}
  [Q_k,\,R_k]=\mathop{\arg\max}_{Q_i\in\mathbb{Q},R_j\in\mathbb{R}} (P(Q_i,\,R_j\,|\,\mathcal{M}^k)).
\label{eq:maxprob}
\end{equation} 
Subsequently, we estimate the optimal state vector and covariance matrices at the step $k$ with $[Q_k,\,R_k]$ and are ready for filtering at the step $k+1$.

The differences between signal and background tracks are expected to be the largest at the two ends, 
because of the features of single-electron track and double-electron track. Therefore for both two ends of track,
an optimal ratio tuned of the principal track length~(20\%) from the linearly interpolated hit points is selected for KFB.
In Fig.~\ref{fig:trajectory_2}, the dashed red rectangle (L-shaped polygon) marks one end of the background (signal) track used for filtering.
The dashed black track represents the \emph{true} trajectory of events in Geant4 simulation.
One can see the red KFB tracks match the MC-truth reasonably well.

\begin{figure}[h]
  \centering 
  \includegraphics[width=.85\textwidth]{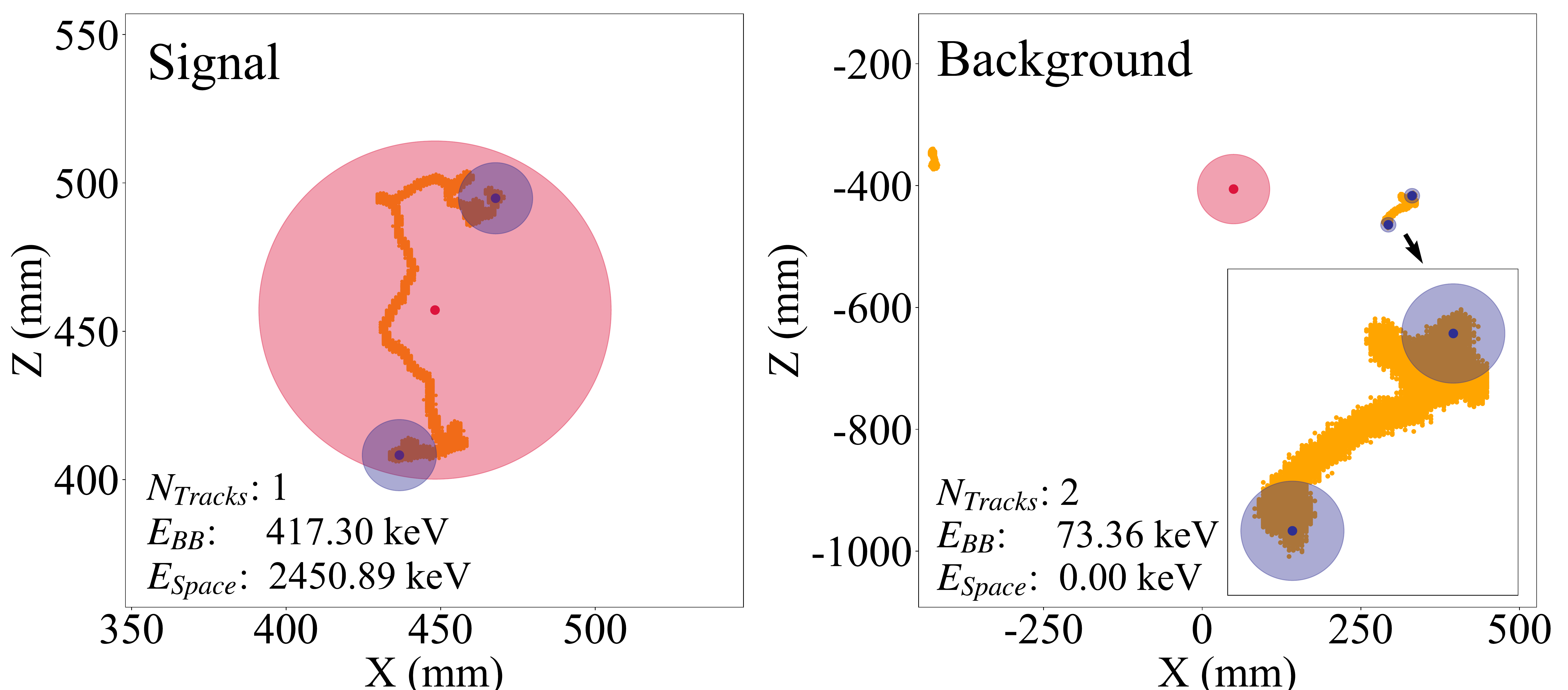}
  \caption{Examples of $0\nu\beta\beta$ signal (left) and $\gamma$ background (right) tracks in the XZ plane for illustrating $N_{Tracks}$, $E_{BB}$, and $E_{Space}$.
  The red circles centered at the charged centers of the events denotes the sphere within which $E_{Space}$ is evaluated. 
  Blob energies are calculated within the blue spheres are the estimated endpoints of the principle tracks.
  $E_{BB}$ shown in the figure are the smaller blob energy at two ends.
  The inset in the right figure shows an enlarged view of the principal track of the background.
  }
  \label{fig:pars_example} 
\end{figure}

\subsection{Track parameters}
\label{sec:Parameters}

Once the event track is reconstructed, we extract six topological parameters for signal identification. 
When the principal track is determined, the total deposited energy of the principal track $E_{p}$ can be calculated.
Both the $0\nu\beta\beta$ signals and the background events may have the multiple tracks, but background events from $\gamma$-rays are more fragmented and hence have smaller $E_p$ in general.
We extract three parameters from the KFB tracks.
With $\theta_{space}^{rms}$ determined in KFB, momenta at the ends of the reconstructed track can also be estimated from Eq.~\ref{eq:theta}.
We define $\hat{P}$ as the larger of momenta at two ends.
BB and Energy loss per unit travel length dE/dx are calculated more precisely along the reconstructed track.
The length of dE/dx and the radius of BB are tuned as 21~mm and 12~mm for the best discrimination power. 

\begin{figure}[tb]
  \centering 
  \includegraphics[width=\textwidth]{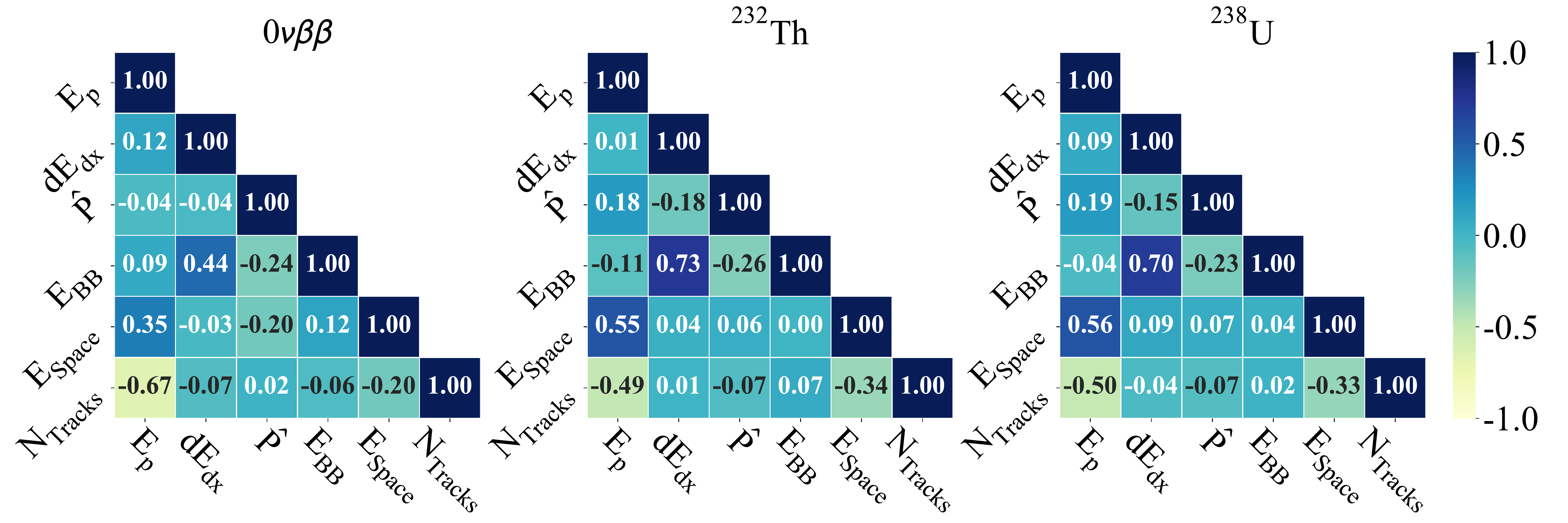}
  \caption{Pearson correlation maps of all the parameters for $0\nu \beta \beta$ signal (left), $^{232}$Th (center), and $^{238}$U (right), respectively. 
  The detector configurations is (1~mm, 3\%).
  }
  \label{fig:Correlation}
\end{figure}

We refer the smaller of dE/dx and BB values at two ends of a track as $dE_{dx}$ and $E_{BB}$ respectively.
However, compared to $E_{BB}$, $dE_{dx}$ is more representative of the Bragg peak of ionization energy loss.
Besides, other two parameters could be calculated accurately together with the subordinate tracks.
The parameter $E_{{Space}}$ defines energy in a unit volume around the energy-weighted center of the event and represents the fragmentation of tracks.
For pixel (strip) readout, the unit volume is a sphere (circle) with a radius of 57~mm, the size of which is optimized for signal and background discrimination.
In addition, the number of tracks $N_{Tracks}$ is also calculated and used to further improve event identification.
Fig.~\ref{fig:pars_example} shows two examples of event tracks to illustrate the definition of $E_{BB}$, $E_{Space}$, and $N_{Trakcs}$. 
Because of the fragmentation of background event tracks, $N_{Tracks}$ and $E_{Space}$ are smaller than these of $0\nu\beta\beta$ signal in general.

The correlation maps of the six parameters for signal and background events under the (1~mm, 3\%) configuration are shown in Fig.~\ref{fig:Correlation}. 
As expected, $dE_{dx}$ and $E_{BB}$ are highly correlated. 
The correlation of background events is about 70\%, while for signal the value is 44\%.
Correlations among $E_{p}$, $E_{space}$, and $N_{Tracks}$ are also observed, because they are all related to the degree of event track dispersion.
However, $E_{p}$, $dE_{dx}$, and $\hat{P}$ show almost no correlation, which demonstrates the validity and the leading role of these three parameters for signal identification.

\begin{figure}[tb]
  \centering 
  \includegraphics[width=0.85\textwidth]{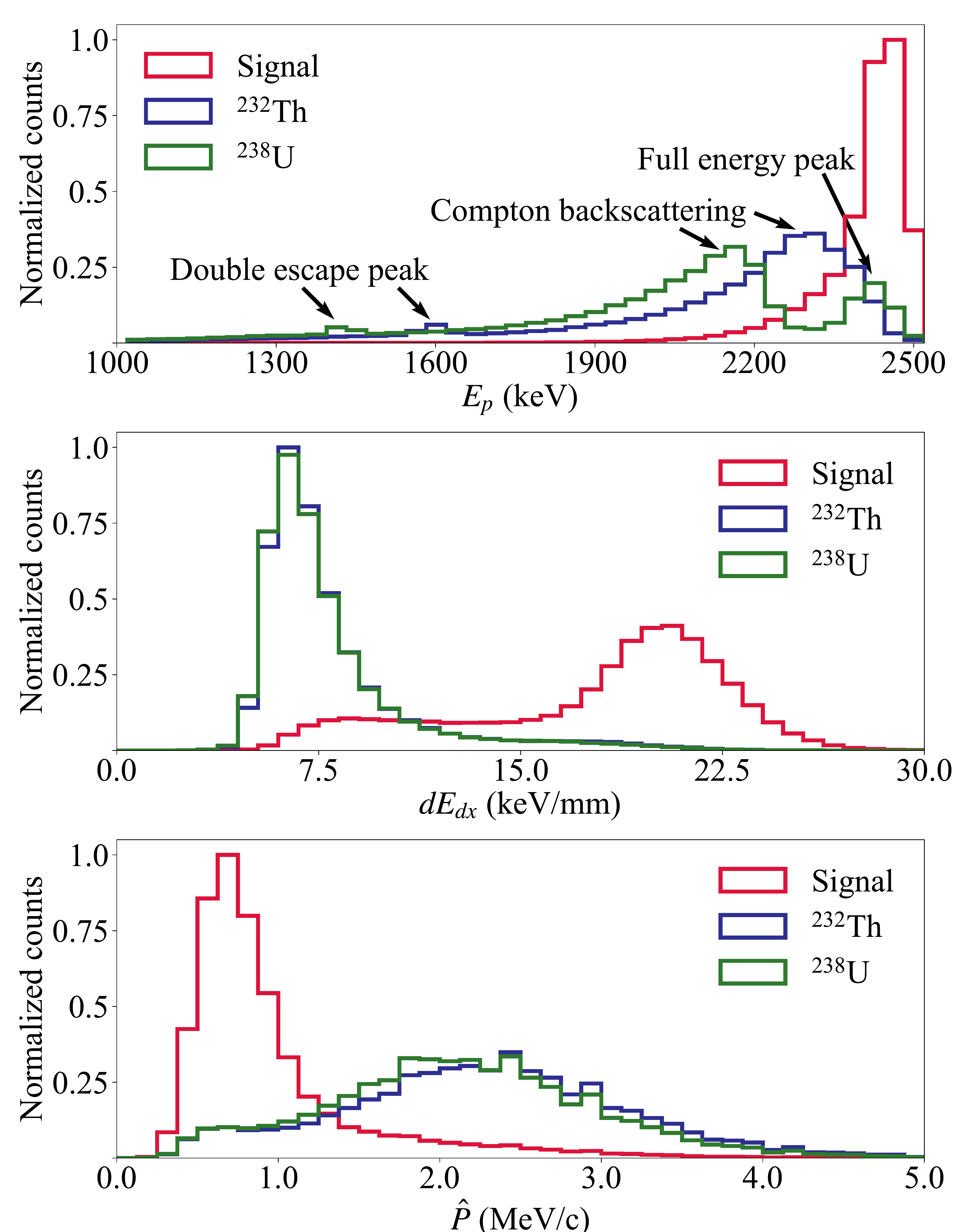}
  \caption{Distributions of the three parameters from the principal track though the KFB workflow, including $E_p$, $dE_{dx}$, and $\hat{P}$, for the (1~mm, 3\%) configuration. 
  $E_{p}$ represents the total deposited energy of the principal track. 
  In the $E_{p}$ distribution of background, double escape peaks, Compton backscattering peaks, and a full absorption peak at 2447~keV are visible and marked.
  $dE_{dx}$ represents the energy loss per unit travel length. 
  The statistical length is 21~mm.
  $\hat{P}$ represents the momenta at the ends of the reconstructed track.
  The Y axis is normalized by the highest value in each plot. 
  }
  \label{fig:Distribution_0} 
\end{figure}

The distributions of $E_{p}$, $dE_{dx}$, and $\hat{P}$ for the (1~mm, 3\%) configuration are shown in Fig.~\ref{fig:Distribution_0}.
In the $E_{p}$ distribution of background, double escape peaks, Compton backscattering peaks, and a full absorption peak at 2447~keV are visible and marked.
The majority of $0\nu\beta\beta$ signals deposit most of the energy in principal tracks and the $E_{p}$ distribution is concentrated within the ROI.
The most effective parameter for the signal and background discrimination is $dE_{dx}$. 
The $dE_{dx}$ of signals is mostly in the range 15 to 30~keV/mm, but background less than 10~keV/mm.
Background events with $dE_{dx}$ more than 10~keV/mm usually deposit a small amount of energy on the principal track, resulting a small $E_p$. 
The distribution of $\hat{P}$ for signal is mainly concentrated in less than 1~MeV/c region, while background is above 1.5~MeV/c.
Because of the wide range of energies in $E_p$, the background events also cover a wide range of $\hat{P}$.
In addition, an excess of background events around 1~MeV/c is due to mis-identification of the principal track. 
When subordinate tracks are too close to the principal one, DBSCAN may group them together and result in smaller $\hat{P}$ values from KFB.
Signal events with $dE_{dx}$ less than 15~keV/mm or $\hat{P}$ over 1~MeV/c are mainly the ones that two electrons share extremely lopsided partition of the Q-value and resemble a background event physically.

\begin{figure}[tb]
  \centering 
  \includegraphics[width=0.85\textwidth]{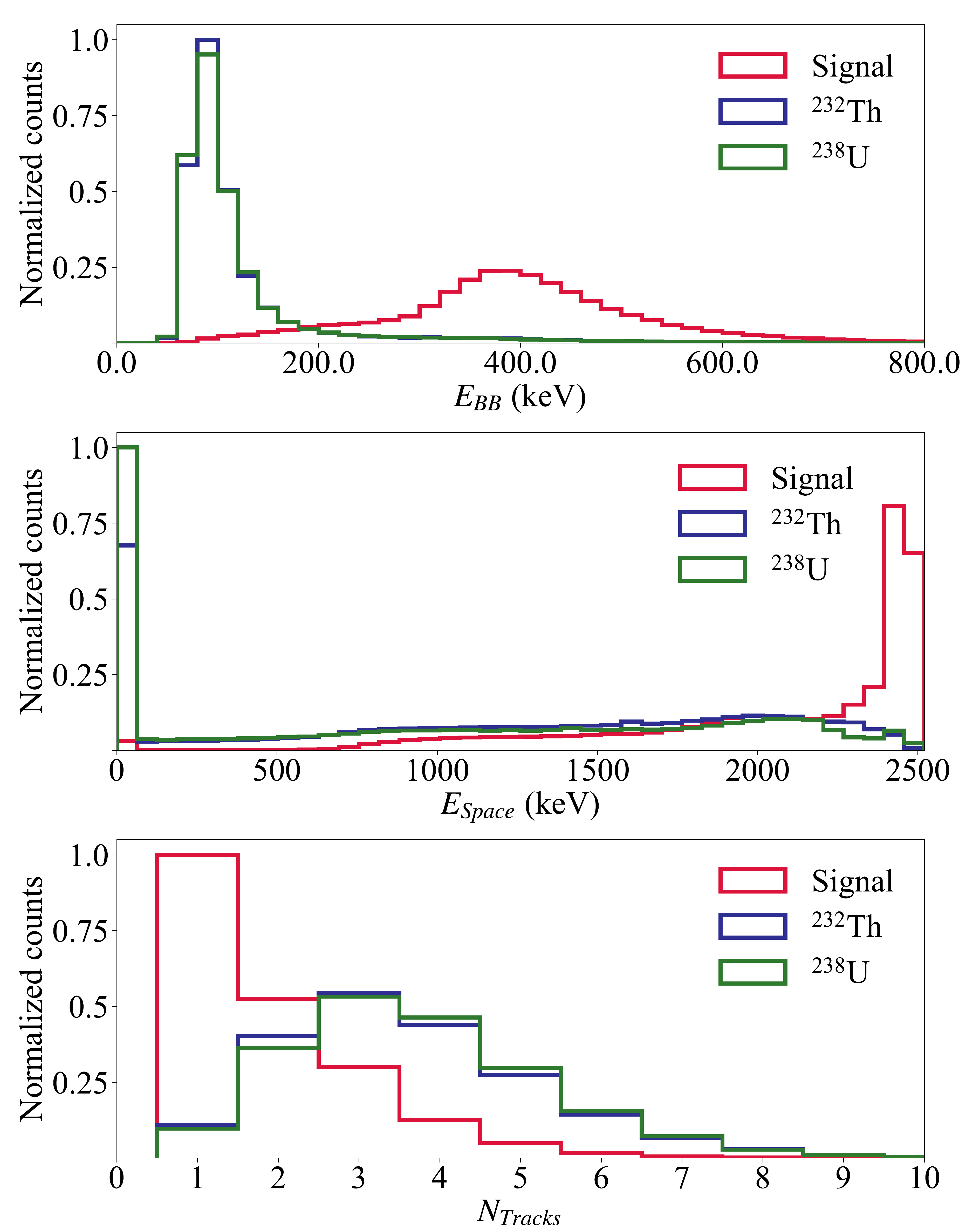}
  \caption{Distributions of $E_{BB}$, $E_{Space}$, and $N_{Tracks}$ for the (1~mm, 3\%) configuration. 
  $E_{BB}$ represents the deposited energy in Bragg blob with radius 12~mm. 
  $E_{Space}$ represents the energy in a unit volume (57~mm) around the energy-weighted center of the event. 
  $N_{Tracks}$ represents the total number of the event track. 
  The Y axis is normalized by the highest value in each plot. 
  }
  \label{fig:Distribution_1} 
\end{figure}

The distributions of $E_{BB}$, $E_{Space}$, and $N_{Tracks}$  for the (1~mm, 3\%) configuration are shown in Fig.~\ref{fig:Distribution_1}.
Compared with $dE_{dx}$ in Fig.~\ref{fig:Distribution_0}, the distribution of $E_{BB}$ is similar but the separation between signal and background is not as distinctive.
The $E_{space}$ distribution of $0\nu\beta\beta$ signals is concentrated within the ROI while these of background events are near zero.
From the $N_{Tracks}$ distribution one can see that about half of $0\nu\beta\beta$ signal has a single track since emitted electrons lose energy via continuous scattering. 
The majority of background events have three or four tracks because $\gamma$ background events deposit energy mainly by Compton scattering at multiple sites. 

\section{Event classification and improvement on sensitivity}
\label{sec:result}

\subsection{Classification significance}

Three parameters from principal track, $E_{p}$, $dE_{dx}$, and $\hat{P}$ are used as rectangular cuts and the results are shown in Table~\ref{tab:Result} for the five configurations.
We define the discrimination significance $\Xi= \left. \epsilon_s \middle/ \sqrt{\epsilon_b} \right.$, where $\epsilon_s$ is the signal efficiency and $\epsilon_b$ the background efficiency after the cuts. 
In the (1~mm, 3\%) configuration, the background rate of $^{232}$Th is suppressed by more than three orders of magnitude while keeping $\epsilon_s$ at about 30\%, and the corresponding $\Xi$ is 12.7.
The discrimination for $^{238}$U chain is less significant because the 2447~keV full absorption peak from $^{214}$Bi overlaps with signals in $E_{p}$. 
In the (1~mm, 6\%) case, $\Xi$ decreases evidently because the Compton backscattering peak of $^{208}$Tl contaminates the wider ROI, similar to what 2447~keV peak does for all configurations.
Compare to the high granularity case with the same energy resolution, $\Xi$ decreases only slightly in the (3~mm, 3\%) configuration because the 1-mm granularity is an over-kill for diffused tracks in a large TPC.
Under the (3~mm, 1\%) configuration, $\Xi$ is up to 13.8 (8.3) for $^{232}$Th~($^{238}$U), thanks to better determination of $E_{p}$.
For the strip-readout scheme, the discrimination significance $\Xi$ of $^{232}$Th and $^{238}$U is reduced to 7.3 and 4.6 respectively.

The three parameters are also used in a Boosted Decision Trees (BDT) based on the Toolkit for MultiVariate Analysis in ROOT~\cite{Hocker:2007ht}.
BDT classifies signal and background with complicated contours in a multi-dimensional parameter space established by training data.
Compared with the rectangular cuts, the improvement of $\Xi$ is smaller than 5\% for all configurations with pixel readout, which demonstrates the orthogonality of the three parameters.
For the strip readout configuration, the tracks are reconstructed in two two-dimensional planes independently and two sets of $dE_{dx}$ and $\hat{P}$ are obtained.
The BDT cuts get more effective and increase $\Xi$ by approximately 20\%.

When $E_{Space}$, $N_{Track}$, and $E_{BB}$ are added as input for BDT as well, we observe additional 20\% improvement in discrimination power approximately for all configurations.
The best $\Xi$ we achieve is 17.4 (10.3) for $^{232}$Th ($^{238}$U) in the (3~mm, 1\%) configuration. 
Table~\ref{tab:Result} shows the results of BDT cuts with all the input parameters.

\begin{table*}
	\centering
	\renewcommand\arraystretch{1.}

	\begin{tabular*}{\hsize}{@{}@{\extracolsep{\fill}}ccccccccccccccc@{}}
		\toprule[0.25mm]
		\multicolumn{3}{c}{\multirow{3}{*}{\large Configurations}} & \multicolumn{6}{c}{\large Rectangular Cuts} \\
		\Xcline{4-9}{0.6pt}
		\multicolumn{3}{c}{} & \multicolumn{3}{c}{$^{232}$Th} & \multicolumn{3}{c}{$^{238}$U} \\
	
		\Xcline{4-6}{0.6pt} \Xcline{7-9}{0.6pt}
		\multicolumn{3}{c}{} & \multicolumn{1}{c}{\large $\epsilon_s$} & {\large $\epsilon_b$} & {\large $\Xi$} & {\large $\epsilon_s$} & {\large $\epsilon_b$} & {\large $\Xi$} \\
		\midrule[0.25mm]
		\multicolumn{3}{c}{(1~mm, 3\%)} & \multicolumn{1}{c}{0.30} & $5.6 \times 10^{-4}$ & 12.7 & 
					0.53 & $4.6 \times 10^{-3}$ & 7.8 \\
		\multicolumn{3}{c}{(1~mm, 6\%)} & \multicolumn{1}{c}{0.34} & $2.3 \times 10^{-3}$ & 7.1 &
					0.47 & $3.7 \times 10^{-3}$ & 7.7 \\
		\multicolumn{3}{c}{(3~mm, 3\%)} & \multicolumn{1}{c}{0.25} & $3.9 \times 10^{-4}$ & 12.7 &
					0.49 & $4.7 \times 10^{-3}$ & 7.1 \\
		\multicolumn{3}{c}{(3~mm, 1\%)} & \multicolumn{1}{c}{0.36} & $6.8 \times 10^{-3}$ & 13.8 &
		           	0.26 & $9.9 \times 10^{-4}$ & 8.3 \\
		\multicolumn{3}{c}{(3~mm strip, 3\%)} & \multicolumn{1}{c}{0.23} & $1.0 \times 10^{-3}$ & 7.3 &
					0.30 & $4.3 \times 10^{-3}$ & 4.6 \\
		\bottomrule[0.25mm]
	\end{tabular*}
	\begin{tabular*}{\hsize}{@{}@{\extracolsep{\fill}}ccccccccccccccc@{}}
		\toprule[0.25mm]
		\multicolumn{3}{c}{\multirow{3}{*}{\large Configurations}} & \multicolumn{6}{c}{\large BDT Cuts} \\
		\Xcline{4-9}{0.6pt}
		\multicolumn{3}{c}{} & \multicolumn{3}{c}{$^{232}$Th} & \multicolumn{3}{c}{$^{238}$U} \\
	
		\Xcline{4-6}{0.6pt} \Xcline{7-9}{0.6pt}
		\multicolumn{3}{c}{} & \multicolumn{1}{c}{\large $\epsilon_s$} & {\large $\epsilon_b$} & {\large $\Xi$} & {\large $\epsilon_s$} & {\large $\epsilon_b$} & {\large $\Xi$} \\
		\midrule[0.25mm]
		\multicolumn{3}{c}{(1~mm, 3\%)}  & \multicolumn{1}{c}{0.34} & $4.7 \times 10^{-4}$ & 15.7 & 
					0.49 & $2.8 \times 10^{-3}$ & 9.3 \\
		\multicolumn{3}{c}{(1~mm, 6\%)} & \multicolumn{1}{c}{0.35} & $1.2 \times 10^{-3}$ & 10.1 & 
					0.57 & $4.2 \times 10^{-3}$ & 8.8 \\
		\multicolumn{3}{c}{(3~mm, 3\%)} & \multicolumn{1}{c}{0.39} & $6.7 \times 10^{-4}$ & 15.1 &
					0.51 & $3.4 \times 10^{-3}$ & 8.7 \\
		\multicolumn{3}{c}{(3~mm, 1\%)} & \multicolumn{1}{c}{0.50} & $8.2 \times 10^{-4}$ & 17.5 & 
		           	0.40 & $1.5 \times 10^{-3}$ & 10.3 \\
		\multicolumn{3}{c}{(3~mm strip, 3\%)} & \multicolumn{1}{c}{0.32} & $8.3 \times 10^{-4}$ & 11.1 & 
					0.46 & $4.6 \times 10^{-3}$ & 6.8 \\
		\bottomrule[0.25mm]
	\end{tabular*}
	\caption{ 
		Effects of rectangular cuts and BDT cuts on  $\epsilon_s$, $\epsilon_b$, and $\Xi$ in the five Configurations.
		Background events from $^{232}$Th and $^{238}$U are considered separately.
		We define the discrimination significance $\Xi= \left. \epsilon_s \middle/ \sqrt{\epsilon_b} \right.$, where $\epsilon_s$ is the signal efficiency and $\epsilon_b$ the background efficiency after the cuts. 
	}
    \label{tab:Result} 
\end{table*}

\subsection{\texorpdfstring{$0\nu\beta\beta$}{TEXT} search sensitivity study}

To illustrate the effectiveness of event discrimination, we calculate the background levels and $0\nu \beta \beta$ search sensitivity of the PandaX-III experiment with and without topological cuts. 
$^{232}$Th and $^{238}$U sources are considered.
Without any topological cuts, the total count of background level is 152 Count per Year (CPY) in the ROI, which is defined as $0.85\times\rm{FWHM}$ (i.e., twice the standard deviation of a Gaussian peak) around the Q-value.
The $^{232}$Th chain contributes to 74\% of the background events and the $^{238}$U chain 26\%.
The majority of the background are from the acrylic field cage, the copper liner, and the stainless vessel.
Table~\ref{tab:background_level} shows the background CPY in different geometry parts.

\newcommand{\tabincell}[2]{\begin{tabular}{@{}#1@{}}#2\end{tabular}}
\begin{table*}
	\centering
	\renewcommand\arraystretch{1.}
	\setlength{\tabcolsep}{1mm}{
	\begin{tabular*}{\hsize}{@{}@{\extracolsep{\fill}}p{2.8cm}<{\centering}ccccccc@{}}
		\toprule[0.25mm]
		\tabincell{c}{Components} & 
		\tabincell{c}{Lead\\shielding} &
		\tabincell{c}{Vessel\\wall} &
		\tabincell{c}{Copper\\liner} &
		\tabincell{c}{Acrylic\\cage} &
		\tabincell{c}{Cathode} &
		\tabincell{c}{Readout\\plane} \\
		\midrule[0.25mm]
		Material & Lead~\cite{Abgrall_2016} & \tabincell{c}{Stainless\\steel~\cite{Akerib:2015cja}} & Copper~\cite{Chen:2016qcd} & Acrylic~\cite{Boger_2000} & Copper~\cite{Chen:2016qcd} & Copper~\cite{Chen:2016qcd} \\
		\midrule[0.25mm]
		$^{232}$Th (CPY) & 0.32 & 11.47 & 8.13 & 91.94 & 0.34 & 0.04 \\
		$^{238}$U (CPY) & 0.08 & 0.97 & 2.57 & 35.03 & 0.45 & 0.22 \\
		\bottomrule[0.25mm]
	\end{tabular*}
	\caption{ Within ROI, the background rates from the  $^{232}$Th and $^{238}$U decay chains in different geometry parts for the configuration of (3~mm strip, 3\%). 
	The majority of the background are from the acrylic field cage, the copper liner, and the stainless vessel.
	}
    \label{tab:background_level} 
    }
\end{table*}

\begin{figure}[tb]
  \centering 
  \includegraphics[width=0.75\textwidth]{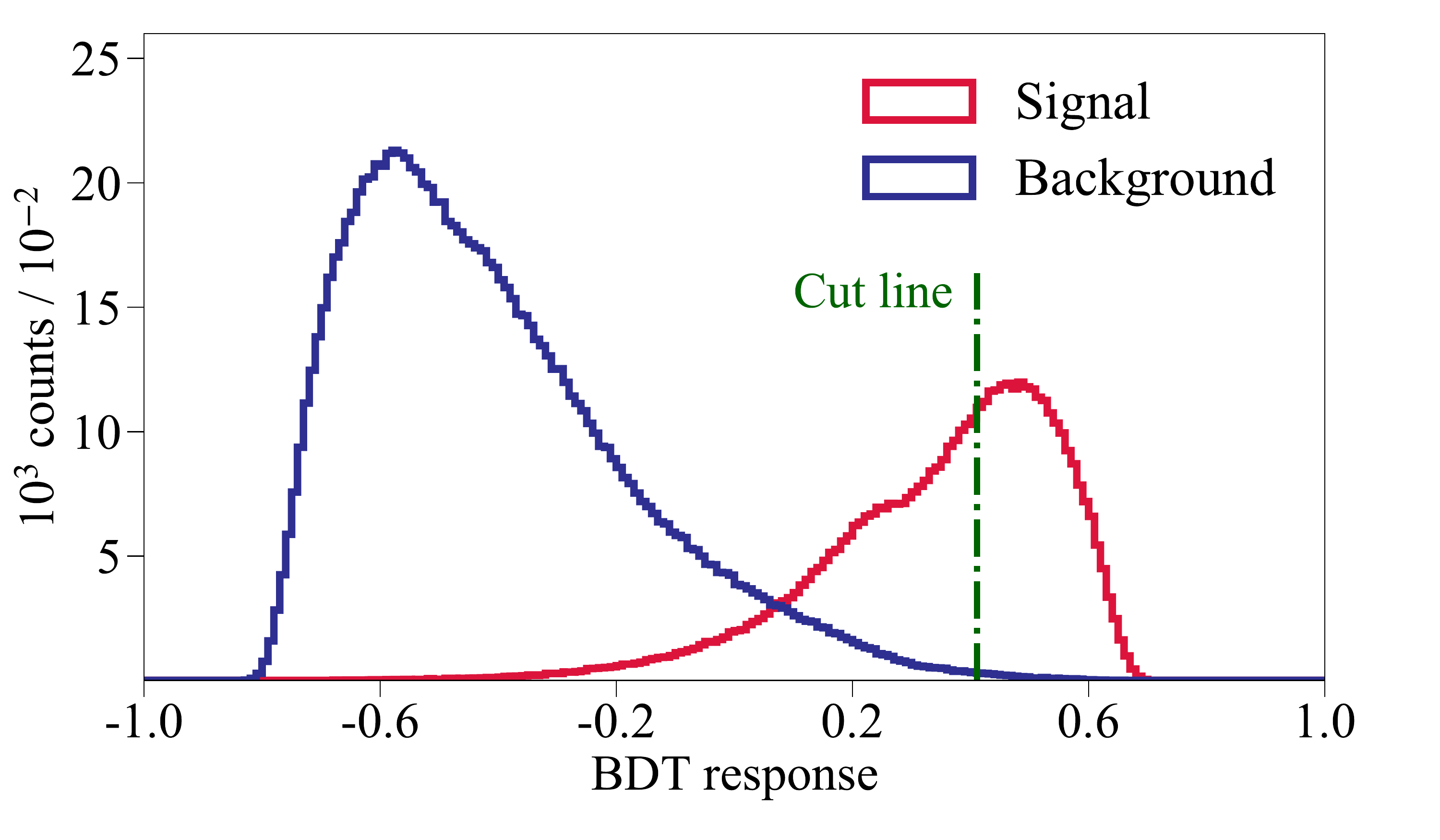}
  \caption{Classification output distribution of $0\nu \beta \beta$ signal (red) and background (blue) with BDT cuts. The optimal cut line (green dash line) is determined for the best $\Xi$.  
  } 
  \label{fig:BDT_Response} 
\end{figure}

\begin{figure}[tb]
  \centering 
  \includegraphics[width=\columnwidth]{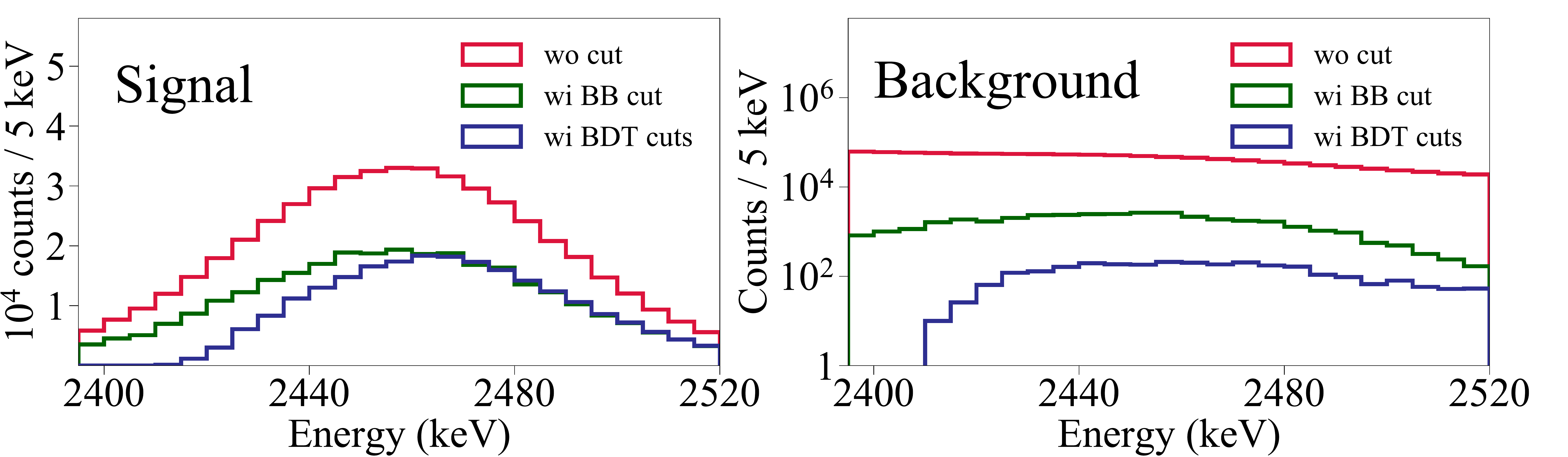}
  \caption{Effects of BB cut and BDT cuts for the signal (left) and background (right). 
  The background count after BDT cuts obviously goes down by about an order of magnitude, compared with that after BB cut.
  In the BDT cuts, Events with energy less than 2410 keV are rejected because of the $E_p$ influence.
  }
  \label{fig:Spectrum} 
\end{figure}

The BDT is re-trained with mixed $^{232}$Th and $^{238}$U backgrounds and cut criteria re-adjusted to maximize the search sensitivity of $0\nu \beta \beta$.
Fig.~\ref{fig:Spectrum} shows the signal and background spectra without cuts, with BB cut, and with BDT cuts respectively. 
We can see the background count after BDT cuts goes down by an order of magnitude, compared with that after BB cut.
Fig.~\ref{fig:BDT_Response} shows the classification results of BDT cuts. The cut line for the best $\Xi$ is shown as the green dash line.
Within ROI, the optimized BDT cuts suppress background by $3.2 \times 10^{-3}$, resulting a background of 0.49~CPY.
The corresponding signal efficiency of topological cuts is 50\% and 
the overall efficiency is 35\%. 
Compared with the design target~\cite{Chen:2016qcd}, the background rate is 10.4 times lower, thanks to the BDT cuts based on KFB tracks.
The background rate is 3.1 times smaller than the one with the original feature cuts developed in Ref.~\cite{Galan:2019ake} and meanwhile the signal efficiency is 1.5 times higher.
Assuming null results, the exclusion limits are established following a method outlined in Ref.~\cite{Alessandria:2011rc} to take into account of Poissonian fluctuation of small background values.
The exclusion sensitivity of PandaX-III is $2.7 \times 10^{26}$~yr (90\% CL) with 5 year live time, a factor of 2.7 or 2.4 improvement comparing to the two cases.

\section{Conclusion and discussion}
\label{sec:conclusion}

In summary, we present a Kalman Filter based reconstruction workflow for meandering tracks of MeV-scale electrons traveling in a high pressure gas medium.
With Monte Carlo simulated data, we apply the technique to $0\nu \beta \beta$ searches, and demonstrate a significant improvement in signal and background discrimination. 
Five different configurations of readout schemes and energy resolutions for a  gaseous TPC are compared in terms of discrimination power. 
Impact of the spatial granularity of 1~mm and 3~mm is marginal, partially because track width due to electron diffusion is nominally larger than detector granularity.
On the other hand, deterioration of energy resolution can negatively affect the identification of background events from $^{232}$Th and lower the discrimination power. 
For the best configuration of 3~mm granularity and 1\% resolution, cuts based on the BDT classifier can suppress $^{232}$Th ($^{238}$U) background by a factor of $8.3 \times 10^{-4}$ ($1.5 \times 10^{-3}$), while keeping 50\% (40\%) of the signals.
In our simulated detector, a background rate of 0.11 CPY is achieved with a signal efficiency of 40\%.
Equivalently less than one background event is expected in 5 years, pushing the search towards background-free regime.  

With a strip-readout scheme as the one currently used in PandaX-III, we can still suppress the background by a factor of more than 300 while keeping at least 50\% of the signal with topological cuts from our improved track reconstruction.
After the traditional calorimeter cuts and BDT cuts, the background rate is 0.49 CPY for PandaX-III, and the corresponding half-life sensitivity to $0\nu \beta \beta$ is $2.7 \times 10^{26}$~yr (90\% CL) for an exposure of five years.
Compared to the previously published topological cuts~\cite{Galan:2019ake}, our new approach improves the search sensitivity by a factor of 2.4.

Further improvement of the workflow can be expected by Extended Kalman filter (EKF)~\cite{Fruhwirth:178627}.
EKF may take into account of ionization energy loss together with Coulomb scattering processes and estimate the momenta of the particle more accurately.
One can reconstruct the entire tracks with momentum information along the track, which would be a powerful background suppression tool for background-free searches of $^{136}$Xe $0\nu \beta \beta$.

Broader application of the reconstruction of meandering track with KFB is warranted.
We could estimate the measurement noise in KFB to study the diffusion effect of the tracks, which can help determine the position in drift direction for the PandaX-III experiment. 
The track reconstruction can also possibly locate the $0\nu \beta \beta$ vertex, which facilitates the tagging of barium ions in a gaseous TPC~\cite{McDonald:2017izm}.
In addition, EKF can be used to determine the energy and angular distributions of the electrons emitted from $2\nu \beta \beta$ and $0\nu \beta \beta$ processes to help understand the decay mechanisms. 
We also envision possible application of the KFB track reconstruction in determining directions of meandering charged particles tracks in general. 
For example, the direction of MeV-GeV scale Compton or pair-produced electron tracks of $\gamma$-ray imaging telescopes~\cite{Mizumura:2013tda,Gros:2017wyj} can be better reconstructed with our approach to improve the angular resolution.

\acknowledgments

This work is supported by the grant from the National Key R\&D Program of China (No. 2016YFA0400300) and the grants from National Natural Sciences Foundation of China (No. 11775142 and No. 11905127). This work is supported in part by the Chinese Academy of Sciences Center for Excellence in Particle Physics (CCEPP).

\bibliographystyle{JHEP}
\bibliography{manuscript}

\providecommand{\href}[2]{#2}\begingroup\raggedright\begin{thebibliography}{10}

\bibitem{Majorana:1937vz}
E.~Majorana, \emph{{Teoria simmetrica dell\textquoteright{}elettrone e del
  positrone}}, \href{https://doi.org/10.1007/BF02961314}{\emph{Nuovo Cim.}
  {\bfseries 14} (1937) 171}.

\bibitem{Avignone:2007fu}
F.T.~Avignone, III, S.R.~Elliott and J.~Engel, \emph{{Double Beta Decay,
  Majorana Neutrinos, and Neutrino Mass}},
  \href{https://doi.org/10.1103/RevModPhys.80.481}{\emph{Rev. Mod. Phys.}
  {\bfseries 80} (2008) 481} [\href{https://arxiv.org/abs/0708.1033}{{\ttfamily
  0708.1033}}].

\bibitem{Dolinski:2019nrj}
M.J.~Dolinski, A.W.P.~Poon and W.~Rodejohann, \emph{{Neutrinoless Double-Beta
  Decay: Status and Prospects}},
  \href{https://doi.org/10.1146/annurev-nucl-101918-023407}{\emph{Ann. Rev.
  Nucl. Part. Sci.} {\bfseries 69} (2019) 219}
  [\href{https://arxiv.org/abs/1902.04097}{{\ttfamily 1902.04097}}].

\bibitem{KamLAND-Zen:2016pfg}
{\scshape KamLAND-Zen} collaboration, \emph{{Search for Majorana Neutrinos near
  the Inverted Mass Hierarchy Region with KamLAND-Zen}},
  \href{https://doi.org/10.1103/PhysRevLett.117.082503}{\emph{Phys. Rev. Lett.}
  {\bfseries 117} (2016) 082503}
  [\href{https://arxiv.org/abs/1605.02889}{{\ttfamily 1605.02889}}].

\bibitem{Agostini:2020xta}
{\scshape GERDA} collaboration, \emph{{Final Results of GERDA on the Search for
  Neutrinoless Double-$\beta$ Decay}},
  \href{https://doi.org/10.1103/PhysRevLett.125.252502}{\emph{Phys. Rev. Lett.}
  {\bfseries 125} (2020) 252502}
  [\href{https://arxiv.org/abs/2009.06079}{{\ttfamily 2009.06079}}].

\bibitem{Adams:2019jhp}
{\scshape CUORE} collaboration, \emph{{Improved Limit on Neutrinoless
  Double-Beta Decay in $^{130}$Te with CUORE}},
  \href{https://doi.org/10.1103/PhysRevLett.124.122501}{\emph{Phys. Rev. Lett.}
  {\bfseries 124} (2020) 122501}
  [\href{https://arxiv.org/abs/1912.10966}{{\ttfamily 1912.10966}}].

\bibitem{Luscher:1998sd}
R.~Luscher et~al., \emph{{Search for beta beta decay in Xe-136: New results
  from the Gotthard experiment}},
  \href{https://doi.org/10.1016/S0370-2693(98)00906-X}{\emph{Phys. Lett. B}
  {\bfseries 434} (1998) 407}.

\bibitem{Rogers:2018lle}
{\scshape NEXT} collaboration, \emph{{High Voltage Insulation and Gas
  Absorption of Polymers in High Pressure Argon and Xenon Gases}},
  \href{https://doi.org/10.1088/1748-0221/13/10/P10002}{\emph{JINST} {\bfseries
  13} (2018) P10002} [\href{https://arxiv.org/abs/1804.04116}{{\ttfamily
  1804.04116}}].

\bibitem{Ferrario:2015kta}
{\scshape NEXT} collaboration, \emph{{First proof of topological signature in
  the high pressure xenon gas TPC with electroluminescence amplification for
  the NEXT experiment}},
  \href{https://doi.org/10.1007/JHEP01(2016)104}{\emph{JHEP} {\bfseries 01}
  (2016) 104} [\href{https://arxiv.org/abs/1507.05902}{{\ttfamily
  1507.05902}}].

\bibitem{Chen:2016qcd}
X.~Chen et~al., \emph{{PandaX-III: Searching for neutrinoless double beta decay
  with high pressure$^{136}$Xe gas time projection chambers}},
  \href{https://doi.org/10.1007/s11433-017-9028-0}{\emph{Sci. China Phys. Mech.
  Astron.} {\bfseries 60} (2017) 061011}
  [\href{https://arxiv.org/abs/1610.08883}{{\ttfamily 1610.08883}}].

\bibitem{Qiao:2018edn}
H.~Qiao, C.~Lu, X.~Chen, K.~Han, X.~Ji and S.~Wang, \emph{{Signal-background
  discrimination with convolutional neural networks in the PandaX-III
  experiment using MC simulation}},
  \href{https://doi.org/10.1007/s11433-018-9233-5}{\emph{Sci. China Phys. Mech.
  Astron.} {\bfseries 61} (2018) 101007}
  [\href{https://arxiv.org/abs/1802.03489}{{\ttfamily 1802.03489}}].

\bibitem{Galan:2019ake}
J.~Galan et~al., \emph{{Topological background discrimination in the PandaX-III
  neutrinoless double beta decay experiment}},
  \href{https://doi.org/10.1088/1361-6471/ab4dbe}{\emph{J. Phys. G} {\bfseries
  47} (2020) 045108} [\href{https://arxiv.org/abs/1903.03979}{{\ttfamily
  1903.03979}}].

\bibitem{doi:10.1002/acs.2369}
P.~Matisko and V.~Havlena, \emph{Noise covariance estimation for kalman filter
  tuning using bayesian approach and monte carlo},
  \href{https://doi.org/10.1002/acs.2369}{\emph{Int. J. Adapt. Control Signal
  Process.} {\bfseries 27} (2013) 957}.

\bibitem{Frosini:2017ftq}
M.~Frosini and D.~Bernard, \emph{{Charged particle tracking without magnetic
  field: optimal measurement of track momentum by a Bayesian analysis of the
  multiple measurements of deflections due to multiple scattering}},
  \href{https://doi.org/10.1016/j.nima.2017.06.030}{\emph{Nucl. Instrum. Meth.
  A} {\bfseries 867} (2017) 182}
  [\href{https://arxiv.org/abs/1706.05863}{{\ttfamily 1706.05863}}].

\bibitem{Agostinelli:2002hh}
{\scshape GEANT4} collaboration, \emph{{GEANT4--a simulation toolkit}},
  \href{https://doi.org/10.1016/S0168-9002(03)01368-8}{\emph{Nucl. Instrum.
  Meth. A} {\bfseries 506} (2003) 250}.

\bibitem{Wang:2020owr}
S.~Wang, \emph{{The TPC detector of PandaX-III Neutrinoless Double Beta Decay
  experiment}},
  \href{https://doi.org/10.1088/1748-0221/15/03/C03052}{\emph{JINST} {\bfseries
  15} (2020) C03052} [\href{https://arxiv.org/abs/2001.01356}{{\ttfamily
  2001.01356}}].

\bibitem{Xie:2020xmd}
C.~Xie, K.~Ni, K.~Han and S.~Wang, \emph{{Enhanced search sensitivity to double
  beta decay of $^{136}$Xe to excited states with topological signatures}},
  \href{https://doi.org/10.1007/s11433-020-1693-6}{\emph{Sci. China Phys. Mech.
  Astron.} {\bfseries 64} (2021) }
  [\href{https://arxiv.org/abs/2012.04552}{{\ttfamily 2012.04552}}].

\bibitem{Ponkratenko:2000um}
O.A.~Ponkratenko, V.I.~Tretyak and Y.G.~Zdesenko, \emph{{The Event generator
  DECAY4 for simulation of double beta processes and decay of radioactive
  nuclei}}, \href{https://doi.org/10.1134/1.855784}{\emph{Phys. Atom. Nucl.}
  {\bfseries 63} (2000) 1282}
  [\href{https://arxiv.org/abs/nucl-ex/0104018}{{\ttfamily nucl-ex/0104018}}].

\bibitem{Adam:2003kg}
W.~Adam, R.~Fruhwirth, A.~Strandlie and T.~Todorov, \emph{{Reconstruction of
  electrons with the Gaussian sum filter in the CMS tracker at LHC}},
  \href{https://doi.org/10.1088/0954-3899/31/9/N01}{\emph{eConf} {\bfseries
  C0303241} (2003) TULT009}
  [\href{https://arxiv.org/abs/physics/0306087}{{\ttfamily physics/0306087}}].

\bibitem{Piacquadio:2008zza}
G.~Piacquadio and C.~Weiser, \emph{{A new inclusive secondary vertex algorithm
  for b-jet tagging in ATLAS}},
  \href{https://doi.org/10.1088/1742-6596/119/3/032032}{\emph{J. Phys. Conf.
  Ser.} {\bfseries 119} (2008) 032032}.

\bibitem{Chatterjee:2014vta}
A.~Chatterjee, K.K.~Meghna, K.~Rawat, T.~Thakore, V.~Bhatnagar, R.~Gandhi
  et~al., \emph{{A Simulations Study of the Muon Response of the Iron
  Calorimeter Detector at the India-based Neutrino Observatory}},
  \href{https://doi.org/10.1088/1748-0221/9/07/P07001}{\emph{JINST} {\bfseries
  9} (2014) P07001} [\href{https://arxiv.org/abs/1405.7243}{{\ttfamily
  1405.7243}}].

\bibitem{Fruhwirth:178627}
R.~Frühwirth, \emph{Application of kalman filtering to track and vertex
  fitting}, \href{https://doi.org/10.1016/0168-9002(87)90887-4}{\emph{Nucl.
  Instrum. Methods Phys. Res., A} {\bfseries 262} (1987) 444}.

\bibitem{10.5555/3001460.3001507}
M.~Ester, H.-P.~Kriegel, J.~Sander and X.~Xu, \emph{A density-based algorithm
  for discovering clusters in large spatial databases with noise},  in
  \emph{Proceedings of the Second International Conference on Knowledge
  Discovery and Data Mining}, KDD'96, p.~226–231, AAAI Press, 1996.

\bibitem{10.1145/235968.233324}
T.~Zhang, R.~Ramakrishnan and M.~Livny, \emph{Birch: An efficient data
  clustering method for very large databases},
  \href{https://doi.org/10.1145/235968.233324}{\emph{SIGMOD Rec.} {\bfseries
  25} (1996) 103–114}.

\bibitem{dorigo1997ant}
M.~Dorigo and L.M.~Gambardella, \emph{Ant colony system: a cooperative learning
  approach to the traveling salesman problem},
  \href{https://doi.org/10.1109/4235.585892}{\emph{IEEE Trans. Evol. Comput.}
  {\bfseries 1} (1997) 53}.

\bibitem{10.1115/1.3662552}
R.E.~Kalman, \emph{{A New Approach to Linear Filtering and Prediction
  Problems}}, \href{https://doi.org/10.1115/1.3662552}{\emph{J. Basic Eng.}
  {\bfseries 82} (1960) 35}.

\bibitem{Innes:1992ge}
W.R.~Innes, \emph{{Some formulas for estimating tracking errors}},
  \href{https://doi.org/10.1016/0168-9002(93)90942-B}{\emph{Nucl. Instrum.
  Meth. A} {\bfseries 329} (1993) 238}.

\bibitem{Patrignani:2016xqp}
{\scshape Particle Data Group} collaboration, \emph{{Review of Particle
  Physics}}, \href{https://doi.org/10.1088/1674-1137/40/10/100001}{\emph{Chin.
  Phys. C} {\bfseries 40} (2016) 100001}.

\bibitem{Hocker:2007ht}
A.~Hocker et~al., \emph{{TMVA - Toolkit for Multivariate Data Analysis}},
  \href{https://arxiv.org/abs/physics/0703039}{{\ttfamily physics/0703039}}.

\bibitem{Abgrall_2016}
N.~Abgrall and et~al., \emph{The majorana demonstrator radioassay program},
  \href{https://doi.org/10.1016/j.nima.2016.04.070}{\emph{Nucl. Instrum. Meth.
  A} {\bfseries 828} (2016) 22–36}.

\bibitem{Akerib:2015cja}
{\scshape LZ} collaboration, \emph{{LUX-ZEPLIN (LZ) Conceptual Design Report}},
   \href{https://arxiv.org/abs/1509.02910}{{\ttfamily 1509.02910}}.

\bibitem{Boger_2000}
J.e.a.~Boger, \emph{The sudbury neutrino observatory},
  \href{https://doi.org/10.1016/s0168-9002(99)01469-2}{\emph{Nucl. Instrum.
  Meth. A} {\bfseries 449} (2000) 172–207}.

\bibitem{Alessandria:2011rc}
{\scshape CUORE} collaboration, \emph{{Sensitivity of CUORE to Neutrinoless
  Double-Beta Decay}},  \href{https://arxiv.org/abs/1109.0494}{{\ttfamily
  1109.0494}}.

\bibitem{McDonald:2017izm}
A.D.~McDonald et~al., \emph{{Demonstration of Single Barium Ion Sensitivity for
  Neutrinoless Double Beta Decay using Single Molecule Fluorescence Imaging}},
  \href{https://doi.org/10.1103/PhysRevLett.120.132504}{\emph{Phys. Rev. Lett.}
  {\bfseries 120} (2018) 132504}
  [\href{https://arxiv.org/abs/1711.04782}{{\ttfamily 1711.04782}}].

\bibitem{Mizumura:2013tda}
Y.~Mizumura et~al., \emph{{Development of a 30 cm-cube Electron-Tracking
  Compton Camera for the SMILE-II Experiment}},
  \href{https://doi.org/10.1088/1748-0221/9/05/C05045}{\emph{JINST} {\bfseries
  9} (2014) C05045} [\href{https://arxiv.org/abs/1312.0438}{{\ttfamily
  1312.0438}}].

\bibitem{Gros:2017wyj}
P.~Gros et~al., \emph{{Performance measurement of HARPO: A time projection
  chamber as a gamma-ray telescope and polarimeter}},
  \href{https://doi.org/10.1016/j.astropartphys.2017.10.008}{\emph{Astropart.
  Phys.} {\bfseries 97} (2018) 10}
  [\href{https://arxiv.org/abs/1706.06483}{{\ttfamily 1706.06483}}].

\end{thebibliography}\endgroup

\end{document}